\documentclass[twoside]{article}
\usepackage{qic}
\usepackage{algorithm}
\usepackage{algorithmic}
\usepackage{amsmath}
\usepackage{amsthm}
\usepackage{amssymb}
\usepackage{hyperref}

\hypersetup{
    colorlinks=true,
    linkcolor=blue,
    urlcolor=cyan,
    citecolor=green
}

\usepackage{tikz}
\usepackage{lipsum}
\usepackage{float}
\usepackage{physics}

\newtheorem{theorem}{Theorem}[subsection]
\newtheorem{lemma}[theorem]{Lemma}

\newtheorem{definition}[theorem]{Definition}
\newtheorem{example}[theorem]{Example}

\newtheorem{remark}{Remark}[subsection]

\numberwithin{equation}{section}

\begin{document}
\setlength{\textheight}{8.0truein}   

\runninghead{Molchanov's Formula and Quantum Walks: A Probabilistic Approach}
            {Hoang Vu}

\normalsize\textlineskip
\thispagestyle{empty}

\vspace*{0.88truein}


\fpage{1}

\centerline{\bf
Molchanov's Formula and Quantum Walks: A Probabilistic Approach}
\vspace*{0.035truein}
\vspace*{0.37truein}
\centerline{\footnotesize
Hoang Vu}
\vspace*{0.015truein}
\centerline{\footnotesize\it Department of Statistics and Applied Probability, University of California, Santa Barbara}
\baselineskip=10pt
\centerline{\footnotesize\it Santa Barbara, California 93106, United States}
\vspace*{10pt}

\vspace*{0.21truein}

\abstracts{This paper establishes a robust link between quantum dynamics and classical one by deriving probabilistic representations for both continuous-time and discrete-time quantum walks (QWs). We first adapt Molchanov’s formula, originally employed in the study of Schrödinger operators on the lattice $\mathbb Z^d$, to characterize the evolution of continuous-time QWs. Extending this framework, we develop a probabilistic methodology to represent the discrete-time QWs on an infinite integer line, bypassing the locality constraints that typically inhibit direct extensions of Molchanov’s approach. The validity of our representation is empirically confirmed through a benchmark analysis of the Hadamard walk, demonstrating high fidelity with traditional unitary evolution. Our results suggest that this probabilistic lens offers a powerful alternative for simulating high-dimensional quantum walks and provides new analytical pathways for investigating quantum systems via classical stochastic processes.
}{}{}

\vspace*{10pt}

\vspace*{3pt}

\vspace*{1pt}\textlineskip    
\section{Introduction}        
Quantum walks (QWs) serve as a powerful generalization of classical random walks, providing a fundamental framework for quantum information processing and algorithm design. Broadly categorized into discrete-time (coined) and continuous-time variants, QWs have been the subject of rigorous study since the seminal work of Gudder \cite{gudder} and the subsequent exploration of quantum lattice gas automata by Meyer \cite{meyer}. The unique ballistic spreading of the discrete-time Hadamard walk—first detailed by Nayak and Vishwanath \cite{nayak} and Ambainis et al. \cite{amb}—diverges sharply from the diffusion patterns governed by the classical Central Limit Theorem. This departure was formally codified by Konno \cite{konno, konno2}, who established a distinct weak limit theorem for one-dimensional lattices, a result later generalized by Grimmett et al. \cite{grimmett}. Despite these advances, extending such limit theorems to multi-dimensional manifolds remains an analytical challenge that has not been fully resolved to date.\\

Historically, the analytical toolkit for QWs has been dominated by combinatorial methods \cite{konno} and Fourier analysis within functional analytic frameworks \cite{grimmett}. Conversely, a purely probabilistic approach has remained underdeveloped. This scarcity is largely due to the fundamental nature of the QW: it is a deterministic, unitary evolution rather than a stochastic process. However, recent literature \cite{konno3, mon, yama} has begun to suggest that viewing QWs through the lens of probability theory reveals deep, previously hidden structural symmetries between quantum and classical dynamics. By employing a probabilistic representation, we can uncover these latent relationships and leverage classical stochastic tools for quantum systems.\\

In this paper, we bridge this gap by adapting Molchanov’s formula—a classical probabilistic tool originally developed for Schrödinger operators on the lattice $\mathbb Z^d$ \cite{carmona}—to the study of quantum dynamics. In Section \ref{sec2}, we establish a formal mapping between Molchanov’s representation and the continuous-time quantum walk, defined via the solution to the Schrödinger equation \cite{ch}. While locality constraints \cite{po} preclude a direct extension of Molchanov’s formula to the discrete-time case, we introduce a novel alternative methodology in Section \ref{sec3}. This method yields a robust probabilistic representation for discrete-time QWs on an integer line driven by arbitrary coin matrices. In parallel with our works, the authors in a working paper cite{ji} obtained a different representation; however, it lacks empirical validation and contains several analytical inconsistences.\\

Finally, in Section \ref{sec4}, we propose and implement efficient algorithms to simulate quantum walks based on our derived probabilistic formulas. We verify our theoretical results through a benchmark analysis of the Hadamard walk. By framing the quantum walk as a probabilistic structure, we provide a new vantage point for investigating high-dimensional discrete-time walks, offering a scalable pathway for Monte Carlo simulations and the eventual derivation of multi-dimensional weak limit theorems in future research.  

\section{Molchanov's Probabilistic Formula for The Continuous-time Quantum Walk}{\label{sec2}}
\noindent

We will first define the continuous-time quantum walk:

\vspace*{12pt}
\noindent
\begin{definition}{\label{def1}}
     Let $(X_t)_{t\geq 0}$ be the continuous-time Markov chain with the probability transition matrix $P$, and the jump times of the chain is denoted by the Poisson process $(N_t)_{t\geq 0}$ with parameter $\lambda>0$. The continuous-time quantum walk $Q$ on is determined by the unitary evolution operator $U(t)=e^{i\lambda Pt}$ such that the quantum state $\Psi$ at time $t\geq 0$ is:
 \begin{align*}
     \ket{\Psi(t)}=U(t)\ket{\Psi(0)}.
 \end{align*}
 In the other words, it is the solution of the following Schrodinger equation:
 \begin{align}
     i\frac{\partial\Psi}{\partial t}=-\lambda P\Psi.\label{eq1}
 \end{align}
\end{definition}
\vspace*{12pt}
\noindent

The Molchanov formula is established in 1981, and has been used to study Schrodinger operation on lattice $\mathbb Z^d$ (see e.g. \cite{carmona}). However, using Definition \ref{def1}, we can modify it to obtain the probabilistic formula for the continuous-time quantum walk on an infinite integer line $\mathbb Z$. The Molchanov's representation of such a walk is stated in the following theorem:

\vspace*{12pt}
\noindent
\begin{theorem}
     A continuous-time quantum walk in Definition \ref{def1} admits the following probabilistic representation:
 \begin{align}
     \Psi(t,x)=e^{\lambda t}\mathbb E\Big[i^{N_t}\Psi(0,X_t)\Big],\label{eq2}
 \end{align}
where $\Psi(.)$ represent the probability amplitude of the walk.
\end{theorem}
\vspace*{12pt}
\noindent

\begin{proof}
    It is sufficient to show that from Equation \eqref{eq2} we can obtain Equation \eqref{eq1}. Indeed, we have:
    \begin{align*}
        \Psi(t+\Delta t,x)=e^{\lambda (t+\Delta t)}\mathbb E\Big[i^{N_{t+\Delta_t}}\Psi(0,X_{t+\Delta t})\Big]
    \end{align*}
    Applying the law of total expectation and condition on $N_{\Delta t}$, we obtain:
    \begin{align*}
        \Psi(t+\Delta_t,x)=e^{\lambda (t+\Delta_t)}\mathbb E\Big[i^{N_{t+\Delta_t}}\Psi(0,X_{t+\Delta_t})\Big|N_{\Delta_t}=0\Big]\cdot \mathbb P[N_{\Delta_t}=0]\\+e^{\lambda (t+\Delta_t)}\mathbb E\Big[i^{N_{t+\Delta_t}}\Psi(0,X_{t+\Delta_t})\Big|N_{\Delta_t}=1\Big]\cdot \mathbb P[N_{\Delta_t}=1] \\+O(\Delta_t^2)
    \end{align*}
    \begin{align*}
        =e^{\lambda (t+\Delta_t)}e^{-\lambda \Delta_t}\mathbb E\Big[i^{N_{t+\Delta_t}}\Psi(0,X_{t+\Delta_t})\Big|N_{\Delta_t}=0\Big]\\+e^{\lambda (t+\Delta_t)}e^{-\lambda \Delta_t}(\lambda \Delta_t)\mathbb E\Big[i^{N_{t+\Delta_t}}\Psi(0,X_{t+\Delta_t})\Big|N_{\Delta_t}=1\Big] \\+O(\Delta_t^2).
    \end{align*}
    Now, using time-homogenity, we obtains:
    \begin{align*}
        \Psi(t+\Delta_t,x)&=e^{\lambda t}\mathbb E\Big[i^{N_{t}}\Psi(0,X_{t})\Big]+e^{\lambda t}(\lambda \Delta_t)(iP)\mathbb E\Big[i^{N_{t}}\Psi(0,X_{t})\Big] +O(\Delta_t^2)\\
        &=\Psi(t,x)+\Delta_t (i\lambda P)\Psi(x,t) + O(\Delta_t^2).
    \end{align*}
    Thus, we have:
    \begin{align*}
        \frac{\Psi(t+\Delta_t,x)-\Psi(t,x)}{\Delta_t}=(i\lambda P)\Psi(x,t) + O(\Delta_t^2).
    \end{align*}
    Taking the limit and let $\Delta_t\to 0$ completes the proof.    
\end{proof}

  One can attempt to derive a discrete-time version of the Molchanov formula. For example, we can define a sequence of $n$ i.i.d Poisson random variables $N_{j, j=1,...,n}$ with parameter $\lambda>0$, and easily show that the probability amplitude evolution after $n$-steps satisfies the following probabilistic representation:
  \begin{align}
      \Psi(n,x)=e^{\lambda n}\mathbb E\Big[i^{\sum_{j=1}^nN_j}\Psi(0,X_n)\Big].\label{eq3}
  \end{align}
However, the discrete-time quantum walk here is not well-defined due to locality (see e.g. \cite{po}). Hence, we need to find a different approach to get the probabilistic representation for discrete-time quantum walk via coin model. Nevertheless, we will soon see that the correct representation of discrete-time quantum walk is not much different from Equation \eqref{eq3}.

\section{A Probabilistic Representation of Discrete-time Quantum Walk}{\label{sec3}}

Let us define the discrete-time quantum walk via the Hilbert space $\mathcal H$ such that 
\begin{align*}
    \mathcal H=\ell^2(\mathbb Z,\mathbb C^2)=\bigg\{\Psi:\mathbb Z\rightarrow \mathbb C^2\bigg |\sum_{x\in \mathbb Z}||\Psi(x)||^2_{\mathbb C^2}<\infty\bigg\},
\end{align*}
where $\mathbb Z$ corresponds to the integer lattice of walker's position space, $\mathbb C^2$ corresponds to the complex coin space, and $\Psi$ is the quantum states.\\ 

We denote the Banach space of bounded operators in $\mathcal H$ by $\mathcal L(\mathcal H)$ and its closed subgroup of unitary operators by $\mathcal U(\mathcal H)$. The standard orthonormal basis of the coin space is $\{-1, 1\}$, which are defined as:
\begin{align*}
    \ket{-1}:=\begin{pmatrix}1\\0\end{pmatrix};\quad \quad 
    \ket{1}:=\begin{pmatrix}0\\1\end{pmatrix}.
\end{align*}

Then, the discrete-time quantum walk is defined as follows:

\vspace*{12pt}
\noindent
\begin{definition}{\label{def0}}
     A random quantum walk $Q$ under the Hilbert space $\mathcal H=\ell^2(\mathbb Z) \otimes \ell^2(\mathbb C^2)$, where the position space denoted by $\ell^2(\mathbb Z)=\text{Span}\{|x\rangle, x\in\mathbb Z\}$ and the coin space denoted by $\ell^2(\mathbb C^2)=\text{Span}\{\ket{y}, y=\pm 1\}$, is determined by the unitary evolution operator $U\in \mathcal L(\mathcal H)$:
    \begin{align}
        U=S\cdot \Big(\sum_{x\in \mathbb Z}|x\rangle\langle x|\otimes C(x)\Big),
    \end{align}
    where $S$ is the shift operator such that
    \begin{align}
        S\ket{x}\otimes\ket{y}=\ket{x+y}\otimes\ket{y},
    \end{align}
    and $C(x)\in \mathcal U(\ell^2(\mathbb C^2))$ is the quantum coin. 
\end{definition}
\vspace*{12pt}
\noindent

Note that any coin matrix $C\in \mathcal U(\ell^2(\mathbb C^2))$ can be written in the following form via the Euler angle decomposition:
\begin{align}
C=e^{i\lambda_1\sigma_3}e^{i\lambda_2\sigma_2}e^{i\lambda_3\sigma_3},\label{eq33}
\end{align}
where $\lambda_j\in (0,2\pi), j=1,2,3$; $\sigma_2$, and $\sigma_3$ are Pauli matrix $Y$ and $Z$ respectively, and are defined as follows:
\begin{align*}
    \sigma_2=\begin{pmatrix}
        0 &-i\\
        i & 0
    \end{pmatrix}, \quad \quad \sigma_3=\begin{pmatrix}
        1 &0 \\
        0 &-1
    \end{pmatrix}.
\end{align*}

This motivates us to look at the probabilistic representation of the quantum walk associated with the coin matrix $\sigma_2$ and $\sigma_3$ first before deriving the formula for the walk with general coin.

\subsection{A Formula for The Pauli Coins}
Let us first consider the coin $C=e^{i\lambda \sigma_2}$, we have:

\vspace*{12pt}
\noindent
\begin{lemma}{\label{lem4}}
    The probability amplitude evolution of a discrete-time quantum walk driven by the homogeneous coin $C=e^{i\lambda \sigma_2}$ follows:
         \begin{align}
        \Psi_n(x,y)=\sum_{k_1,k_2,...,k_n\in \mathbb N}i^{\sum_{j=1}^nk_j+y_j\cdot\frac{1-(-1)^{k_j}}{2}}\frac{\lambda^{\sum_{j=1}^nk_j}}{k_1!k_2!...k_n!}\Psi_0(x_n,y_n),\label{eq6}
      \end{align}  
    where $x_n:=x_0-\sum_{j=0}^{n-1}y_j$, $y_n:=(-1)^{k_n}y_{n-1}$ for $n\geq 1$ with $(x_0,y_0)=(x,y)$.  
\end{lemma}

\vspace*{12pt}
\noindent
\begin{proof}   
    For any state $\Psi$ of the walk, we have:
     \begin{align*}
        U\Psi&=U\sum_{\substack{x\in\mathbb Z\\y\in\{\pm 1\}}}\Psi(x,y)\ket{x}\ket{y}\\
        &=S\cdot(I\otimes C)\sum_{\substack{x\in\mathbb Z\\y\in\{\pm 1\}}}\Psi(x,y)\ket{x}\ket{y}\\
        &=\sum_{\substack{x\in\mathbb Z\\y\in\{\pm 1\}}}\Psi(x,y)S\ket{x}e^{i\lambda \sigma_2}\ket{y} \\
        &=\sum_{\substack{x\in\mathbb Z\\y\in\{\pm 1\}}}\Psi(x,y)\sum_{k\in \mathbb N}S\ket{x}\frac{(i\lambda)^k}{k!}\sigma_2^k\ket{y}\\
        &=\sum_{\substack{x\in\mathbb Z\\y\in\{\pm 1\}\\k\in \mathbb N}}\Psi(x,y)i^{k}\frac{\lambda^k}{k!}i^{y\cdot\frac{1-(-1)^k}{2}}\ket{x+(-1)^ky}\ket{(-1)^ky}\\
        &=\sum_{\substack{x\in\mathbb Z\\y\in\{\pm 1\}\\k\in \mathbb N}}i^{k+y\cdot\frac{1-(-1)^k}{2}}\frac{\lambda^k}{k!}\Psi(x-y,(-1)^ky)\ket{x}\ket{y}.
    \end{align*}   
   
   This implies that 
   \begin{align}
       (U\Psi)(x,y)=\sum_{k\in\mathbb N}i^{k+y\cdot\frac{1-(-1)^k}{2}}\frac{\lambda^k}{k!}\Psi(x-y,(-1)^ky).
   \end{align}
  
    Hence, the evolution after $n-$steps yields the probability amplitude:
    \begin{align}
      \Psi_n(x,y)=\sum_{k_1,k_2,...,k_n\in \mathbb N}i^{\sum_{j=1}^nk_j+y_j\cdot\frac{1-(-1)^{k_j}}{2}}\frac{\lambda^{\sum_{j=1}^nk_j}}{k_1!k_2!...k_n!}\Psi_0(x_n,y_n).
   \end{align}
   This completes our proof. 
\end{proof}
\vspace*{12pt}
\noindent

We introduce the following classical process to formulate our probabilistic representation:

\vspace*{12pt}
\noindent
\begin{definition}{\label{def4}}
    Let $N_1,N_2,...N_n$ be i.i.d Poisson random variables with parameter $\lambda \in (0,2\pi)$, we have:
    \begin{align*}
    S_0&=0, \quad \quad S_n=\sum_{j=1}^nN_j  \quad \quad (n\geq 1),\\
    Y_0&=y, \quad \quad Y_n=(-1)^{S_n}\bigg(Y_0+\frac{a_{0,c}(Y_0)}{2}\bigg)-\frac{a_{0,c}(Y_0)}{2}(-1)^{S_n(c+1)} \quad \quad (n\geq 1),\\
    X_0&=x, \quad \quad X_n=X_{n-1}-Y_{n-1}=X_0-\sum_{j=0}^{n-1}Y_j \quad \quad (n\geq 1),
    \end{align*}
    where $a_{0,c}(Y_0)$ is a deterministic function of $y$ and $c$, and $c$ is a given fixed constant.
\end{definition}

\vspace*{12pt}
\noindent
\begin{remark}
    The defintion of $Y_n$ could be simpler here, but to keep it consistently with future research on high dimensional quantum walks, we insist to keep it in such a form.
\end{remark}
\vspace*{12pt}
\noindent

    This leads to the following representation theorem:
    
\vspace*{12pt}
\noindent
\begin{theorem}
    A discrete-time quantum walk driven by the homogeneous coin $C=e^{i\lambda \sigma_2}$ has the following probabilistic representation:

    \begin{align}
        \Psi_n(x,y)=e^{n\lambda}\mathbb E\Big [i^{S_n+Y_0\cdot\frac{1-(-1)^{S_n}}{2}}\Psi_0(X_n,Y_n)\Big],
    \end{align}
    for $(x,y,n)\in\mathbb Z\times\{\pm1\}\times\mathbb N_0$, with  $\Psi_0(.,.)$ is defined by Equation \eqref{eq00}, and the classical processes $S_n$, $Y_n$, and $X_n$ are defined in Definition \ref{def4} with $c=0$.
\end{theorem}

\vspace*{12pt}
\noindent
\begin{proof}
    From Equation \eqref{eq6} in Lemma \ref{lem4}, apply the Poisson distribution, we have:

    \begin{align*}
         \Psi_n(x_0,y_0)&=\sum_{k_1,...,k_n\in\mathbb N}i^{\sum_{j=1}^n k_j+y_j\cdot\frac{1-(-1)^{k_j}}{2}}\frac{\lambda^{\sum_{j=1}^nk_j}}{k_1!...k_n!}\Psi_0(x_n,y_n)\\
         &=e^{n\lambda}\sum_{k_1,...,k_n\in\mathbb N}i^{\sum_{j=1}^n k_j+y_j\cdot\frac{1-(-1)^{k_j}}{2}}\frac{e^{-\lambda}\lambda^{k_1}...e^{-\lambda}\lambda^{k_n}}{k_1!...k_n!}\Psi_0(x_n,y_n)\\
         &=e^{n\lambda}\mathbb E\Big [i^{S_n+Y_0\cdot\frac{1-(-1)^{S_n}}{2}}\Psi_0(X_n,Y_n)\Big],
    \end{align*}
    for $x_0=x$, and $y_0=y$. This completes our proof.
\end{proof}
\vspace*{12pt}
\noindent

Now consider the coin $C=e^{i\lambda \sigma_3}$, we have:

\vspace*{12pt}
\noindent
\begin{lemma}{\label{lem7}}
    The probability amplitude evolution of a discrete-time quantum walk driven by the homogeneous coin $C=e^{i\lambda \sigma_3}$ follows:
         \begin{align}
        \Psi_n(x,y)=\sum_{k_1,k_2,...,k_n\in \mathbb N}i^{y_0\sum_{j=1}^nk_j}\frac{\lambda^{\sum_{j=1}^nk_j}}{k_1!k_2!...k_n!}\Psi_0(x_n,y_n),\label{eq7}
      \end{align}  
    where $x_n:=x_0-ny_0$, $y_n:=y_0$ for $n\geq 1$ with $(x_0,y_0)=(x,y)$.  
\end{lemma}

\vspace*{12pt}
\noindent
\begin{proof}    
    For any state $\Psi$ of the walk, we have:
     \begin{align*}
        U\Psi&=U\sum_{\substack{x\in\mathbb Z\\y\in\{\pm 1\}}}\Psi(x,y)\ket{x}\ket{y}\\
         &=S\cdot(I\otimes C)\sum_{\substack{x\in\mathbb Z\\y\in\{\pm 1\}}}\Psi(x,y)\ket{x}\ket{y}\\
        &=\sum_{\substack{x\in\mathbb Z\\y\in\{\pm 1\}}}\Psi(x,y)S\ket{x}e^{i\lambda \sigma_3}\ket{y} \\
         &=\sum_{\substack{x\in\mathbb Z\\y\in\{\pm 1\}}}\Psi(x,y)\sum_{k\in \mathbb N}S\ket{x}\frac{(i\lambda)^k}{k!}\sigma_3^k\ket{y}\\
        &=\sum_{\substack{x\in\mathbb Z\\y\in\{\pm 1\}\\k\in \mathbb N}}\Psi(x,y)i^{k}\frac{\lambda^k}{k!}i^{k(y-1)}\ket{x+y}\ket{y}\\
        &=\sum_{\substack{x\in\mathbb Z\\y\in\{\pm 1\}\\k\in \mathbb N}}i^{ky}\frac{\lambda^k}{k!}\Psi(x-y,y)\ket{x}\ket{y}.
    \end{align*}   
   
   This implies that 
   \begin{align}
       (U\Psi)(x,y)=\sum_{k\in\mathbb N}i^{ky}\frac{\lambda^k}{k!}\Psi(x-y,y).
   \end{align}
  
    Hence, the evolution after $n-$steps yields the probability amplitude:

    \begin{align}
      \Psi_n(x,y)=\sum_{k_1,k_2,...,k_n\in \mathbb N}i^{y_0\sum_{j=1}^nk_j}\frac{\lambda^{\sum_{j=1}^nk_j}}{k_1!k_2!...k_n!}\Psi_0(x-ny_0,y_0).
   \end{align}
    
   This completes our proof. 
\end{proof}
\vspace*{12pt}
\noindent

    This leads to the following representation theorem:

\vspace*{12pt}
\noindent
\begin{theorem}{\label{theo6}}
    A discrete-time quantum walk driven by the homogeneous coin $C=e^{i\lambda \sigma_3}$ has the following representation:
    \begin{align}
        \Psi_n(x,y)=e^{in\lambda y_0}\Psi_0(x_0-ny_0,y_0),
    \end{align}
    for $(x,y,n)\in\mathbb Z\times\{\pm1\}\times\mathbb N_0$ with $(x_0,y_0)=(x,y)$.
\end{theorem}

\vspace*{12pt}
\noindent
\begin{proof}
    From Equation \eqref{eq7} in Lemma \ref{lem7}, apply the Poisson distribution, we have:

    \begin{align*}
         \Psi_n(x_0,y_0)&=\sum_{k_1,...,k_n\in\mathbb N}i^{y_0\sum_{j=1}^n k_j}\frac{\lambda^{\sum_{j=1}^nk_j}}{k_1!...k_n!}\Psi_0(x_n,y_n)\\
         &=e^{n\lambda}\sum_{k_1,...,k_n\in\mathbb N}i^{y_0\sum_{j=1}^n k_j}\frac{e^{-\lambda}\lambda^{k_1}...e^{-\lambda}\lambda^{k_n}}{k_1!...k_n!}\Psi_0(x_0-ny_0,y_0)\\
         &=e^{n\lambda}\Psi_0(x_0-ny_0,y_0)\mathbb E\Big [i^{y_0S_n}\Big]\\
         &=e^{n\lambda}\Psi_0(x_0-ny_0,y_0)\mathbb E\Big [e^{i\frac{\pi}{2}{y_0S_n}}\Big]\\
         &=e^{in\lambda y_0}\Psi_0(x_0-ny_0,y_0),
    \end{align*}

     for $x_0=x$, and $y_0=y$, and where in the last equation we use the characteristic function formula for a Poisson random variable. This completes our proof.
\end{proof}
\vspace*{12pt}
\noindent

\subsection{A Formula for The General Coin}
Now, consider the general coin in Equation \eqref{eq33}, $C=e^{i\lambda_1\sigma_3}e^{i\lambda_2\sigma_2}e^{i\lambda_3\sigma_3}$, we have:

\vspace*{12pt}
\noindent
\begin{lemma}{\label{lem8}}
    The probability amplitude evolution of a discrete-time quantum walk driven by the homogeneous coin $C=e^{i\lambda_1\sigma_3}e^{i\lambda_2\sigma_2}e^{i\lambda_3\sigma_3}$ follows:
         \begin{align}
         \Psi_n(x,y)=\sum_{k_1,k_2,...,k_n\in \mathbb N}e^{i\lambda_1\sum_{j=0}^{n-1}y_j}e^{i\lambda_3 \sum_{j=1}^{n}y_j}i^{\sum_{j=1}^nk_j+y_j\cdot\frac{1-(-1)^{k_j}}{2}}\frac{\lambda_2^{\sum_{j=1}^nk_j}}{k_1!k_2!...k_n!}\Psi_0(x_n,y_n),\label{eq8}
      \end{align}  
    where $x_n:=x_0-\sum_{j=0}^{n-1}y_j$, $y_n:=(-1)^{k_n}y_{n-1}$ for $n\geq 1$ with $(x_0,y_0)=(x,y)$.  
\end{lemma}

\vspace*{12pt}
\noindent
\begin{proof}
    Notice that from Lemma \ref{lem7} and Theorem \ref{theo6}, when only applying the coin $e^{i\lambda_.\sigma_3}$ and keeping the site fixed the one step evolution will be:
    \begin{align*}
        \Psi_1(x,y)=e^{i\lambda_.y_0}\Psi_0(x_0,y_0).
    \end{align*}
     
    Now, for any state $\Psi$ of the walk, we have:
     \begin{align*}
        U\Psi&=U\sum_{\substack{x\in\mathbb Z\\y\in\{\pm 1\}}}\Psi(x,y)\ket{x}\ket{y}\\
        &=S\cdot(I\otimes C)\sum_{\substack{x\in\mathbb Z\\y\in\{\pm 1\}}}\Psi(x,y)\ket{x}\ket{y}\\
        &=\sum_{\substack{x\in\mathbb Z\\y\in\{\pm 1\}}}\Psi(x,y)S\ket{x}e^{i\lambda_1\sigma_3}e^{i\lambda_2\sigma_2}e^{i\lambda_3\sigma_3}\ket{y}\\
        &=\sum_{\substack{x\in\mathbb Z\\y\in\{\pm 1\}}}\Psi(x,y)e^{i\lambda_1y}\sum_{k\in \mathbb N}S\ket{x}e^{i\lambda_3(-1)^ky}\frac{(i\lambda_2)^k}{k!}i^{y\cdot\frac{1-(-1)^k}{2}}\ket{(-1)^ky}\\
        &=\sum_{\substack{x\in\mathbb Z\\y\in\{\pm 1\}\\k\in \mathbb N}}\Psi(x,y)e^{i\lambda_1y}e^{i\lambda_3(-1)^ky}\frac{\lambda_2^k}{k!}i^{k+y\cdot\frac{1-(-1)^k}{2}}\ket{x+(-1)^ky}\ket{(-1)^ky}\\
        &=\sum_{\substack{x\in\mathbb Z\\y\in\{\pm 1\}\\k\in \mathbb N}}e^{i\lambda_1y}e^{i\lambda_3(-1)^ky}\frac{\lambda_2^k}{k!}i^{k+y\cdot\frac{1-(-1)^k}{2}}\Psi(x-y,(-1)^ky)\ket{x}\ket{y}.
    \end{align*}   
   
   This implies that 
   \begin{align}
       (U\Psi)(x,y)=\sum_{k\in\mathbb N}e^{i\lambda_1y}e^{i\lambda_3(-1)^ky}\frac{\lambda_2^k}{k!}i^{k+y\cdot\frac{1-(-1)^k}{2}}\Psi(x-y,(-1)^ky).
   \end{align}
  
    Hence, the evolution after $n-$steps yields the probability amplitude:
    \begin{align}
      \Psi_n(x,y)=\sum_{k_1,k_2,...,k_n\in \mathbb N}e^{i\lambda_1\sum_{j=0}^{n-1}y_j}e^{i\lambda_3 \sum_{j=1}^{n}y_j}i^{\sum_{j=1}^nk_j+y_j\cdot\frac{1-(-1)^{k_j}}{2}}\frac{\lambda_2^{\sum_{j=1}^nk_j}}{k_1!k_2!...k_n!}\Psi_0(x_n,y_n).
   \end{align}
   This completes our proof. 
\end{proof}
\vspace*{12pt}
\noindent

This leads to the following representation theorem:
    
\vspace*{12pt}
\noindent
\begin{theorem}{\label{theo8}}
    A discrete-time quantum walk driven by the homogeneous coin $C=e^{i\lambda_1\sigma_3}e^{i\lambda_2\sigma_2}e^{i\lambda_3\sigma_3}$ has the following probabilistic representation:
    \begin{align}
        \Psi_n(x,y)=e^{n\lambda_2}\mathbb E\Big [i^{S_n+Y_0\cdot\frac{1-(-1)^{S_n}}{2}}e^{i\lambda_1(X_0-X_n)}e^{i\lambda_3(X_0-X_n+Y_n)}\Psi_0(X_n,Y_n)\Big],\label{eq9}
    \end{align}
    for $(x,y,n)\in\mathbb Z\times\{\pm1\}\times\mathbb N_0$, with  $\Psi_0(.,.)$ is defined by Equation \eqref{eq00}, and the classical processes $S_n$, $Y_n$, and $X_n$ are defined in Definition \ref{def4} with $c=0$.
\end{theorem}

\vspace*{12pt}
\noindent
\begin{proof}
    From Equation \eqref{eq8} in Lemma \ref{lem8}, apply the Poisson distribution, we have:
    \begin{align*}
         \Psi_n(x_0,y_0)&=\sum_{k_1,k_2,...,k_n\in \mathbb N}e^{i\lambda_1\sum_{j=0}^{n-1}y_j}e^{i\lambda_3 \sum_{j=1}^{n}y_j}i^{\sum_{j=1}^nk_j+y_j\cdot\frac{1-(-1)^{k_j}}{2}}\frac{\lambda_2^{\sum_{j=1}^nk_j}}{k_1!k_2!...k_n!}\Psi_0(x_n,y_n)\\
         &=e^{n\lambda_2}\sum_{k_1,...,k_n\in\mathbb N}e^{i\lambda_1\sum_{j=0}^{n-1}y_j}e^{i\lambda_3 \sum_{j=1}^{n}y_j}i^{\sum_{j=1}^nk_j+y_j\cdot\frac{1-(-1)^{k_j}}{2}}\frac{e^{-\lambda_2}\lambda_2^{k_1}...e^{-\lambda_2}\lambda_2^{k_n}}{k_1!...k_n!}\Psi_0(x_n,y_n)\\
         &=e^{n\lambda_2}\mathbb E\Big [i^{S_n+Y_0\cdot\frac{1-(-1)^{S_n}}{2}}e^{i\lambda_1(X_0-X_n)}e^{i\lambda_3(X_0-X_n+Y_n)}\Psi_0(X_n,Y_n)\Big],
    \end{align*}
    for $x_0=x$, and $y_0=y$. This completes our proof.
\end{proof}  

\vspace*{12pt}
\noindent

\begin{example}\label{ex1}
    Consider the Hadamard walk with the coin matrix 
    $$H=\frac{1}{\sqrt{2}}\begin{pmatrix}
        1&1\\1&-1
    \end{pmatrix},$$ which can also be written in the form:
    \begin{align*}
        H=e^{i\frac{\pi}{2}\sigma_3}e^{i\frac{\pi}{4}\sigma_2}.
    \end{align*}
    According to Theorem \ref{theo8}, its probabilistic representation is
    \begin{align}
         \Psi_n(x,y)=e^{\frac{n\pi}{4}}\mathbb E\Big [i^{S_n+Y_0\cdot\frac{1-(-1)^{S_n}}{2}}e^{i\frac{\pi}{2}(X_0-X_n)}\Psi_0(X_n,Y_n)\Big].
    \end{align}
\end{example}

\section{Empirical Analysis of The Formula}{\label{sec4}}
In this section, we present an efficient algorithm to simulate the discrete-time quantum walk with a general coin via its probabilistic representation in Equation \eqref{eq9}.\\

A general form of the initial state of the quantum walker is given by:
\begin{align*}
    \ket{\Psi_0}=\ket{0}\otimes\Big(\alpha\ket{1}+\beta\ket{-1}\Big),
\end{align*}
where $\alpha\in \mathbb C$,$\beta\in \mathbb C$, and $|\alpha|^2+|\beta|^2=1$ are the probability amplitudes corresponding to the coin state $\ket{1}$ and $\ket{-1}$ respectively at position $x=0$ at time $t=0$. Hence, we can define the functional form of $\Psi_0(.,.)$ inside the expectation in Equation \eqref{eq9} by 
\begin{align}
    \Psi_0(x,y):=\mathbb I_{x=0}\Big(\alpha\cdot\mathbb I_{y=1}+\beta\cdot \mathbb I_{y=-1}\Big).\label{eq00}
\end{align}

From here, we can even rewrite Equation \eqref{eq9} in a more compact form:
\begin{align}
     \Psi_n(x,y)=e^{n\lambda_2}\mathbb E\Big [i^{S_n+y\cdot\frac{1-(-1)^{S_n}}{2}}e^{i\lambda_1 x}e^{i\lambda_3(x+y(-1)^{S_n})}\Psi_0(X_n,Y_n)\Big].
\end{align}

Now, we introduce the algorithm for the quantum walk with a general coin:

\begin{algorithm}[H]
\caption{Simulation of Discrete-time Quantum Walks Via Probabilistic Representation}
\label{alg:Qwalk}
\begin{algorithmic}[1]
    \REQUIRE Total number of iterations $M$, the time of investigation $n$, $\alpha$ and $\beta$ as coefficients of the initial coin state, $\lambda_1$ and $\lambda_3$ as the Euler decompostion parameters, and $\lambda_2$ as the parameter of Poisson distribution.
    \STATE Initialize the arrays $L$ and $R$ to keep the probability amplitudes at each position $x \in (-n,n), x\in \mathbb Z$ for the coin spin $\{1\}$ and $\{-1\}$ respectively.
    \REPEAT
        \STATE Sample a sequence of $N_j$ the number of jumps at time $j=1,2,...,n$ from Poisson distribution with mean $\lambda_2$.
        \STATE Compute the sequence of sums $S_1,\cdots,S_n$, where $S_n=\sum_{j=1}^n N_j$. 
        \STATE Compute $Y_n^{\{1\}}=(-1)^{S_n}$ and $Y_n^{\{-1\}}=(-1)^{S_n+1}$.
        \STATE Update the $R$ array at position $x=\sum_{j=0}^{n-1}(-1)^{S_j}$: 
        $$R[x]+=e^{n\lambda_2}e^{i\lambda_1 x}e^{i\lambda_3(x+(-1)^{S_n})}\cdot\frac{i^{S_n+\frac{1-(-1)^{S_n}}{2}}}{M}\cdot \big(\alpha\cdot\mathbb I_{Y_n^{\{1\}}=1}+\beta\cdot \mathbb I_{Y_n^{\{1\}}=-1}\big)$$. 
        \STATE Update the $L$ array at position $x=-\sum_{j=0}^{n-1}(-1)^{S_j}$: 
        $$L[x]+=e^{n\lambda_2}e^{i\lambda_1 x}e^{i\lambda_3(x-(-1)^{S_n})}\cdot\frac{i^{S_n-\frac{1-(-1)^{S_n}}{2}}}{M}\cdot \big(\alpha\cdot\mathbb I_{Y_n^{\{-1\}}=1}+\beta\cdot \mathbb I_{Y_n^{\{-1\}}=-1}\big)$$.
    \UNTIL M iterations are done

    \RETURN The arrays $L$ and $R$.
\end{algorithmic}  
\end{algorithm}

Now, comeback to Example \ref{ex1}, we will simulate the Hadamard walk via the traditional approach, which acts as a benchmark, and compare it with the simulation obtained from Algorithm 1. Note that, the initial state of the Hadamard walk is given by
$$
    \ket{\Psi_0}=\ket{0}\otimes\Big(\frac{1}{\sqrt{2}}\ket{1}+i\frac{1}{\sqrt{2}}\ket{-1}\Big).
$$ 

\begin{figure}[H]
\vspace*{13pt}
\centerline{\epsfig{file=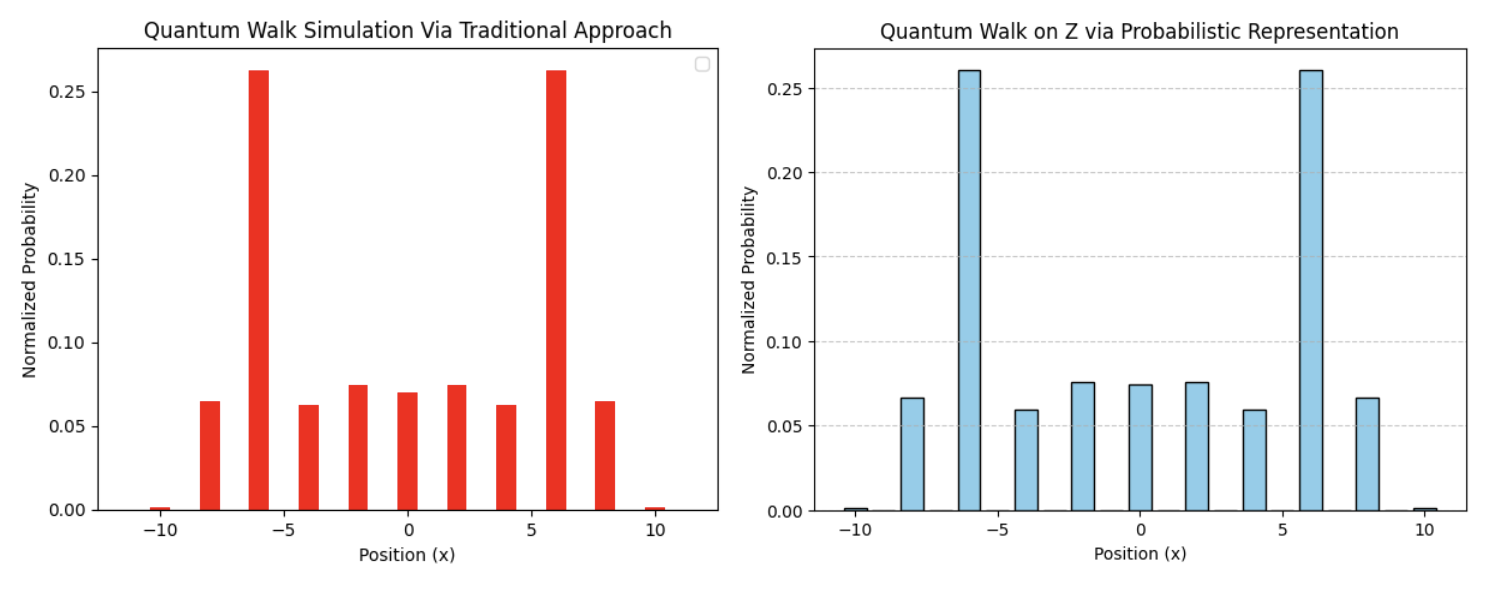, width=15cm}} 
\vspace*{13pt}
\fcaption{\label{fig1} The Hadamard walk's probability distribution for $n=10$, $\alpha=\frac{1}{\sqrt{2}}$, and $\beta=\frac{1}{\sqrt{2}}i$ with the left bar chart illustrating the benchmark method, and the right bar chart illustrating the probabilistic method with the number of iteration $M=5\times10^9$, $\lambda_1=\frac{\pi}{2}$, $\lambda_2=\frac{\pi}{4}$, and $\lambda_3=0$.}
\end{figure}

The numerical simulation results are shown in Figure \ref{fig1}, and confirm the validity of our formula.

\section{Conclusion}
In conclusion, we have explored the intersection of quantum walks and classical stochastic processes by developing a robust probabilistic representation in both continuous and discrete-time cases. While quantum walks are fundamentally deterministic, our work demonstrates that they can be effectively framed through the lens of probability theory, revealing a deeper connection to classical processes than previously emphasized in the literature.\\

We managed to use the Molchanov’s formula, originally a tool for Schrödinger operators, to represent continuous-time quantum walks, and then introduced a methodological framework in Section \ref{sec3} to derive a probabilistic representation for discrete-time quantum walks on an integer line driven by arbitrary coin matrices in $\mathcal U(\mathcal H)$. Furthermore, we demonstrated the practical utility of these theoretical constructions by developing efficient simulation algorithms. Through the specific case of the Hadamard walk, we verified that our probabilistic formulas accurately recover known quantum behaviors, providing a computationally viable alternative to traditional unitary evolution methods.\\

The shift from a functional analysis approach to a probabilistic one opens several promising avenues for future research: our representation provides a potential pathway to overcoming the analytical complexities of multi-dimensional quantum walks, where weak limit theorems remain elusive. In addition, the formulas derived here lay the groundwork for applying variance-reduction techniques and other classical Monte Carlo methods to quantum systems. By mapping quantum amplitudes to probabilistic structures, researchers can possibly identify the specific "quantumness" of a walk in contrast to its classical counterpart.

\nonumsection{References}


\noindent


\begin{thebibliography}{000}
\bibitem{amb}
Ambainis, A., Bach, E., Nayak, A., Vishwanath, A., and Watrous, J. (2001). One-dimensional quantum
walks, Proc. of the 33rd Annual ACM Symposium on Theory of Computing, 37–49.

\bibitem{carmona}
Ren\'e Carmona. Random Schr\"odinger operators. Ecole d’Ete de Probabilites de Saint Flour XIV, pages 1–124,
1984.

\bibitem{child} 
Childs, A. M., Farhi, E., and Gutmann, S. (2002). An example of the diﬀerence between quantum and
classical random walks, Quantum Information Processing, 1, 35–43, quant-ph/0103020.

\bibitem{ch}
Childs, A. M. (2022). Lecture notes on quantum algorithms. University of Maryland. https://www.cs.umd.edu/~amchilds/qa/qa.pdf

\bibitem{grimmett}
Grimmett, G., Janson, S., and Scudo, P. F. (2004). Weak limits for quantum random walks, Phys. Rev. E,
69, 026119, quant-ph/0309135.

\bibitem{gudder}
Gudder, S. P. (1988). Quantum Probability. Academic Press Inc., CA.

\bibitem{konno}
Konno, N. (2002a). Quantum random walks in one dimension, Quantum Information Processing, 1, 345–354, quant-ph/0206053.

\bibitem{konno2}
Konno, N. (2005a). Limit theorem for continuous-time quantum walk on the line, Phys. Rev. E, 72, 026113, quant-ph/0408140.

\bibitem{konno3}
Konno, N., Matsue, K., and Segawa, E. (2023). A crossover between open quantum random walks to quantum walks. Journal of Statistical Physics, 190(12):202.

\bibitem{meyer}
Meyer, D. A. (1996). From quantum cellular automata to quantum lattice gases, J. Statist. Phys., 85, 551–574, quant-ph/9604003.

\bibitem{mon}
Montero, M. (2017). Quantum and random walks as universal generators of probability distributions. Physical Review A, 95(6):062326.

\bibitem{nayak}
Nayak, A., and Vishwanath, A. (2000). Quantum walk on the line, quant-ph/0010117.

\bibitem{po}
Portugal, R. (2018). Quantum walks and search algorithms (2nd ed.). Springer Nature. https://doi.org/10.1007/978-3-319-97813-0

\bibitem{yama}
Yamagami, T., Segawa, E., Chauvet, N., Rohm, A., Horisaki, R., and Naruse, M. (2022). Directivity of quantum walk via its random walk replica. Complexity, 2022(ID 9021583):114.

\end{thebibliography}
\end{document}